\documentstyle[aps]{revtex}

\begin{document}
\input psfig

\title{WORLDSHEET FORMULATION FOR LATTICE STAGGERED FERMIONS}

\author{ J.M. Aroca \\
Departament de Matem\`atiques, Universitat
Polit\`ecnica de Catalunya,\\
Jordi Girona 1-3, Mod C-3 Campus Nord,\\
08034 Barcelona, Spain.}

\author{Hugo Fort and Rodolfo Gambini \\
Instituto de F\'{\i}sica, Facultad de Ciencias, \\
Tristan Narvaja 1674, 11200 Montevideo, Uruguay}

\maketitle

\begin{abstract} 
The worldsheet formulation is introduced for
lattice gauge theories with dynamical fermions.  The partition
function of lattice compact QED with staggered fermions is
expressed as a sum over surfaces with border on self-avoiding
fermionic paths. 
The surfaces correspond to the world sheets of
loop-like pure electric flux excitations and meson-like
configurations (open electric flux tubes carrying matter fields at
their ends).  The proposed formulation does not have the problem of
the additional doubling of the fermion species due to the
discretization of time.  The gauge non-redundancy and the geometric
transparency are two appealing features of this description.  From
the computational point of view, the partition function involves
fewer degrees of freedom than the Kogut-Susskind formulation and
offers an alternative and more economic framework to perform
numerical computations with dynamical fermions.  
\end{abstract}

\newpage

\section{Introduction}

        The problem of handling dynamical 
fermions is still a major challenge
that faces lattice gauge theory at present.  
Monte Carlo techniques have
provided many important results clarifying several 
points about the dynamic,
and recent computations are achieving 10 \% or better 
accuracy in the spectrum
both for heavy quark and light quark system \cite{lm}.  
However, statistical
algorithms are very expensive in computer time, 
and the increased computing
power of the coming generation of machines without 
further theoretical insight
will probably be insufficient in order to definitely 
improve results.
Basically, the difficulty posed by the fermions
stems from the fact that they are represented not by
ordinary numbers but by anti-commuting Grassmann numbers which
cannot be directly simulated numerically.
Since the fermion field appears quadratically in the action,
the usual procedure is to integrate it out producing
the Mattews-Salam determinant. So, the problem of
including dynamical fermions is reduced to one of
evaluating the determinant of a large matrix. This is
a costly task. 

An alternative to tackle this problem is to resort 
to the analytical methods.
These could be divided into two categories: 
strong-coupling expansion and Hamiltonian 
variational-like methods. 
The principal limitation of the former is the difficulty to 
reach the weak-coupling region. 
An exponent of the second group
is the {\em loop} approach \cite{gt},\cite{gt2}. 
The basis of the loop method can be traced 
to the idea of describing gauge
theories explicitly in terms of Wilson 
loops or holonomies \cite{p}-\cite{mami}
since Yang \cite{y} noticed their 
important role for a complete description of
gauge theories.  
The loop Hamiltonian was given  in
terms of two fundamental operators: the 
Wilson Loop operator (the trace of the
holonomy) and the electric field operator 
(its temporal loop derivative).  The
loops replace the information furnished by the 
vector potential (the connection).  A description in terms 
of loops or strings, besides the general advantage of
only involving the gauge invariant physical 
excitations, is appealing because
all the gauge  invariant  operators have a 
simple geometrical meaning when
realized in the loop space.  Last but not least, 
the interest on loops relies
on the fact that it was realized that this formalism 
goes beyond a simple gauge
invariant description and in fact it 
provides a natural geometrical framework
to treat gauge theories and quantum gravity. 
The introduction by Ashtekar
\cite{a} of a new set of variables that cast 
general relativity in the same
language as gauge theories allowed to apply loop 
techniques as a natural
non-perturbative description of Einstein's theory.

        In 1991 the {\em loop} representation was 
        extended in such a way to
include dynamical staggered \cite{sus} fermions : the 
so-called {\em P-representation} \cite{fg}. Roughly the idea
is to add to the closed pure gauge excitations, open ones
corresponding to ``electromesons''.
Afterwards the P-representation was used 
to perform analytical Hamiltonian
calculations, by means of a cluster approximation, providing 
qualitatively good results for the $(2+1)$ \cite{af2d}
and the $(3+1)$ \cite{af3d} cases when compared 
with the standard Lagrangian
numerical simulations in terms of the fields.
The Hamiltonian method has the serious
drawback of the explosive proliferation of 
clusters with the order of the approximation.

       Thus, our goal was to explore another approach: to build
a classical action in terms of strings and knit together 
the transparency and non-redundance
of the string P-formulation and the power of 
the Lagrangian simulations.  The first
step of this program was the introduction 
of new lattice action for pure QED
in terms of closed strings of electric 
flux (loops) \cite{abf}.  In the pure
case the action is written as a sum of integer 
variables attached to the closed
worldsheets of the loop excitations.  
The second step was to include matter
fields into the string
description, with this aim we considered
the simplest gauge theory: the compact 
scalar electrodynamics (SQED)
\cite{abfs}.  In the case of SQED the 
action is expressed in terms of open
and closed surfaces which correspond to world 
sheets of loop-like pure electric
flux excitations and open electric flux tubes 
carrying matter fields at their
ends. The previous two worldsheets actions were 
simulated using the Metropolis
algorithm being the results quite encouraging \cite{abf},
\cite{abfs}.

        Here we show how to 
introduce dynamical fermions in a worldsheet
or Lagrangian description.
This paper is organized as follows. Secction II
is a `bird-eye-view' review of the P-representation
on a hypercubic lattice. We show the realization of the
lattice QED Hamiltonian in the Hilbert space of paths 
$\{ P \}$. In section III we present the Lagrangian
counterpart of the previous P-representation
and we write the partition function in terms of 
worldsheets of the string-like excitations. 
In section IV, by using the transfer matrix procedure, we check  
that we get the  Hamiltonian of section II
from the path integral of section III. Notice that
this enabled us to get the Kogut-Susskind formulation
via the transfer matrix, this 
is an interesting problem which was not properly 
solved.
Finally, section V is devoted to conclusions
and some remarks.

\vspace{1cm}
        
\section{The P-Representation on the lattice}

        The P-representation offers a 
gauge invariant description of physical
states in terms of kets $\mid P >$, 
where $P$ labels a set of connected paths
$P_x^y$ with ends $x$ and $y$ 
\footnote {For a 
more detailed exposition of the P-representation 
and the realization of the different
operators see the ref. \cite{fg}.}. 
In order to make the connection on the lattice 
between the P-representation and the ordinary 
representation, in terms of the
fermion fields $\psi$ and the gauge fields $U_\mu(x)=
\exp [iea A_\mu(x)]$, we need a gauge invariant
object constructed from them. The most natural candidate 
in the continuum is
\begin{equation}
\Phi (P_x^y) = {\psi}^{\dagger} (x) U(P_x^y)\psi (y),
\end{equation}
where $U(P_x^y)=\exp [iea \int_{P} A_\mu dx^\mu]$.
 
The immediate problem we face is that $\Phi$ 
is not purely an object belonging
to the ``configuration'' basis because it 
includes the canonical conjugate
momentum of $\psi$, ${\psi}^{\dagger}$.  
The lattice offers a solution to this problem consisting in
the decomposition of the fermionic degrees 
of freedom. Let us consider the
Hilbert space of kets $\mid {\psi}_u^{\dagger} 
,{\psi}_d , A_{\mu} >$, where
$u$ corresponds to the $up$ part of the Dirac 
spinor and $d$ to the $down$
part. Those kets are well defined in terms 
of ``configuration'' variables (the
canonical conjugate momenta of ${\psi}_d$ and  
${\psi}_u^\dagger$ are
${\psi}_d^\dagger$ and ${\psi}_u$ respectively.  
Then, the internal product of
one of such kets with one of the path dependent 
representation (characterized
by a lattice path $P_x^y$ with ends $x$ and $y$) is given by
$$
\Phi (P_x^y) \equiv <P_{x;i}^{y;j} \mid {\psi}_u^{\dagger} ,
{\psi}_d, A_{\mu} > 
$$
\begin{equation}
= {\psi}_{u;i}^{\dagger} (x) U(P_x^y) {\psi}_{d;j}(y),
\label{eq:Phi}
\end{equation}
where $i$ and $j$ denote a component of the 
spinor $u$ and $d$ respectively.
Thus, it seems that the choice 
of staggered fermions is the natural one 
in order to build the lattice P-representation.  
Therefore, the lattice 
paths $P_x^y$ start in
sites $x$ of a given parity and end in sites 
$y$ with  opposite parity.
The one spinor component at each site 
can be described in terms of the $\chi (x)$ single Grassmann 
fields \cite{sus}. The path creation  
operator $\hat{\Phi}_Q$ in the space of kets $\{ |\, P> \}$
of a path with ends $x$ and $y$ is defined as 
\begin{equation}
\hat{\Phi}_Q= \hat{\chi}^{\dagger} (x) \hat{U}(Q_x^y) 
\hat{\chi}(y).
\label{eq:Phiop}
\end{equation}
Its adjoint operator $\hat{\Phi}_Q^{\dagger}$ acts
in two possible ways \cite{fg}: annhilating the path $Q_x^y$
or joining two existing paths in $|\, P>$ one ending at
$x$ and the other strarting at $y$.

Let us show the realization of the 
QED Hamiltonian in the Hilbert space of kets $|\, P>$. 
This Hamiltonian is given by
\begin{eqnarray}
\hat{H}=(g^2/2)\hat{W} \nonumber \\
\hat{W}=\hat{W}_E +\lambda 
\hat{W}_I+\lambda^2\hat{W}_M
\nonumber \\
\lambda =1/g^2 \nonumber \\
\hat{W}_E=\sum_\ell \hat{E}_\ell^2 \nonumber \\
\hat{W}_I=-\sum_{\ell} (\hat{\Phi}_\ell + 
\hat{\Phi}_\ell^{\dagger})
\nonumber \\
(\hat{\Phi}_\ell^{\dagger}=\eta_n (x)\hat{\chi}^{\dagger}(x)
\hat{U}_n(x)\hat{\chi} ( x+ n) )
\label{eq:H} \\
\eta_{\mbox{\bf e}_i} (x)=(-1)^{x_1+...+x_{i-1}}, \nonumber \\
\eta_{-\mbox{\bf e}_i} (x+a\mbox{\bf e}_i)= 
\eta_{\mbox{\bf e}_i} (x), \nonumber \\
\hat{W}_M=-\sum_p (\hat{U}_p+\hat{U}_p^\dagger), \nonumber
\end{eqnarray}
where  $x$ labels sites, $\ell \equiv (x, n)$ the spatial
links pointing along the spatial unit vector $n$, 
$p \equiv (x, n, n')$ the spatial plaquettes; 
$\hat{E}_\ell$ is the electric field operator, 
which is diagonal in the P-representation
\begin{equation}
\hat{E}_\ell |\, P \, > = N_\ell (P) |\, P \, >,
\label{eq:E}
\end{equation}
where the eigenvalues $N_\ell(P)$ are the number of times that the 
link $\ell$ appears in  the set of paths $P$;
$\hat{U}_p = \prod_{\ell \in p} \hat{U}_\ell$. 
The $\hat{\Phi}_\ell$ are 
``displacement" operators corresponding to the quantity 
defined in (\ref{eq:Phi}) for the case of a one-link
path i.e. $P\equiv \ell$. The realization of the different 
Hamiltonian terms in this representation is
as follows \cite{fg}:

\vspace{3mm}

First, by (\ref{eq:E})
the action of the electric Hamiltonian is given by
\begin{equation}
\hat{W}_E\mid P >=\sum_\ell N_\ell^2(P)\mid P >.
\label{eq:We}
\end{equation}

The interaction term $\hat{W}_I$ can be written as

$$
-\hat{W}_I=\sum_{x_e,n} 
\hat{\Phi}_n(x_e) + \sum_{x_o,n} 
\hat{\Phi}_n(x_o),$$
where the subscripts $e$ and $o$ denote the parity of
the lattice sites. This term
is realized in $\{ |\, P> \}$ as      
\begin{equation}
-\hat{W}_I\mid P >=\sum_{x_e,n}\epsilon (P,\ell_{x_e} )
\mid P\cdot \ell_{x_e} >+
\sum_{x_o,n} \epsilon (P,\ell_{x_o} )\mid P\cdot \ell_{x_o}>
\label{eq:Wi}
\end{equation}
where $\ell_{x}$ is the link starting in $x$ and 
ending  in  $x+na$. For links of even origin, $ \epsilon (P,\ell )$ 
is zero whenever an end of $\ell$ coincides with an end of $P$, 
it is $-1$ when $\Phi_\ell$ ``deletes" the link $\ell \in P$ 
, dividing one connected component
into two, and it is $+1$ in any other case. 
For links of odd origin, $ \epsilon (P,\ell )$ is zero
unless both ends of $\ell$ coincide with two ends in $P$.
In that case, it is $-1$ when $\ell$ joins two disconnected pieces
and it is $+1$ when it closes a connected piece or when it
annhilates a link. 
The different actions of operators
$\hat{\Phi}_\ell$ over path-states $|P(t)>$ 
are schematically summarized in FIG.1.

\begin{center}
\begin{figure}[t]
\hskip 1cm \psfig{figure=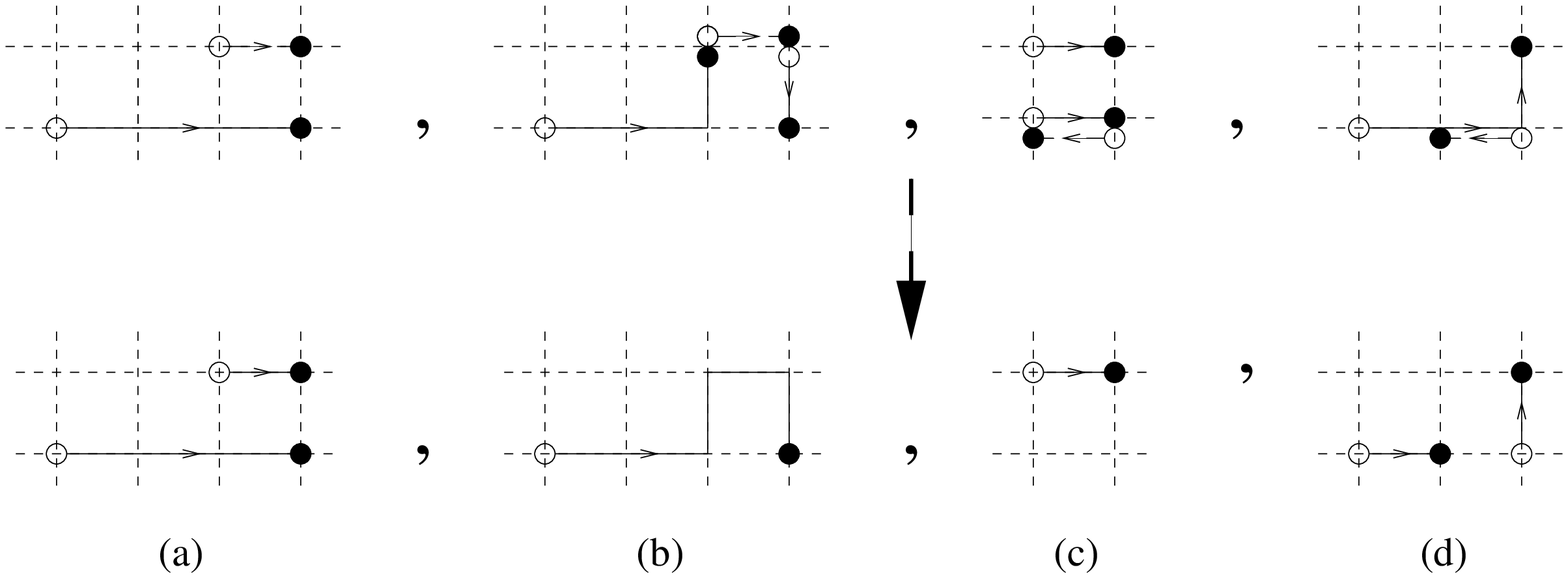,height=5.4cm}
\caption{A summary of the different actions of
operators $\Phi_\ell$ applied over path-states $|P>$.
The link $\ell$ is represented by a dashed bond
in the pictures on the top and the resulting paths 
$|P'>$ are plotted below.}
\label{fig1}
\end{figure}
\end{center}

      Finally $\hat{W}_M$  is the sum of the 
operators $\hat{U}(p)$ and
$\hat{U}^\dagger(p)$ which add plaquettes, and can be written
as\footnote{In such a way that a generic path $P$ is generated 
out from the 0-path state
(strong coupling vacuum) 
by the application of the operator string
$\hat{\chi}^{\dagger} (x)\prod_{\ell \in P_x^y}
\eta_\ell \hat{U}_\ell \hat{\chi}(y)$.}

$$\hat{W}_M= \sum_p \eta_p (\hat{U}_p+\hat{U}_p^\dagger),$$

$\!\!\!\!\!$ where 

$$\eta_p=\prod_{\ell \in p} \eta_\ell=-1,$$

$\!\!\!\!\!$and then
\begin{equation}
\hat{W}_M\mid P>=\sum_p (\mid P \cdot p>+
\mid P \cdot \overline{p}>)
\label{eq:Wm}
\end{equation}
where  $p$  and  $\overline{p}$  respectively  
denote the  clockwise and the
counter-clockwise plaquette contour.

\vspace{1cm}

\section{The Worldsheet or P-action}

       In order to cast the preceding path  
description in the Lagrangian
formalism let us begin by considering the path integral 
for lattice QED with staggered fermions. 
We will show that 
this leads to a surface action which is not 
possible to connect directly with the Hamiltonian (\ref{eq:H})
via the transfer matrix in the Hilbert space $\{|P>\}$.
However, this action will serve as a guide in order to build
the genuine P-action $S_P$. 

For simplicity we choose the 
Villain form of the action, which is given by
$$
Z= \int [ d\chi^\dagger d\chi ] \int [d\theta]\sum_{\{ n_p\}}
\exp\{
-\frac{\beta}{2} \sum_p [ \theta_p +2\pi n_p \,] ^2 
+\frac{1}{2}\sum_\ell a^{(D-1)}
\eta_\ell (\chi^\dagger_r U_\ell \chi_{r+a\hat{\mu}} + H.C.)         
\}$$
\begin{equation}
=\int [ d\chi^\dagger d\chi ] \int [d\theta]\sum_{n}
\exp\{
-\frac{\beta}{2} \|\nabla \theta +2\pi n\|^2 
+\frac{1}{2}\sum_\ell
a^{(D-1)} \eta_\ell (\chi^\dagger_r U_\ell 
\chi_{r+a\hat{\mu}} + H.C.)  
\},
\label{eq:Z}
\end{equation}
where we used in the second line the notations of the 
calculus of differential forms on the lattice
of \cite {g}. In the above expression: 
$D$ is the lattice dimension,
$\beta=\frac{1}{e^2}$,
$\theta$ is a real compact 1-form defined in each link of 
the lattice, $U_\ell=e^{i\theta_\ell}$ and $\chi$ and
$\chi^\dagger$ are Grassmannian 
variables defined on the sites of the lattice,
$\nabla$ is the co-boundary operator --which maps 
$k$-forms into $(k+1)$-forms
--, $n$ are integer 2-forms defined at the 
lattice plaquettes and $\| g \|^2 =
<g,g>=\sum_{c_k} g^2(c_k)$, where $g$ is any 
$k$-form and $c_k$ are the
$k$-cells ($c_0$ sites, $c_1$ links, \ldots ).  
The measure in (\ref{eq:Z}) is
\begin{equation}
[ d \chi^\dagger d\chi ]=\prod_r d \chi^\dagger_r d\chi_r,
\hspace{2cm}
[d\theta]=\prod_\ell\frac{d\theta_\ell}{2\pi}
\label{eq:meas}
\end{equation}

Let us forget for the moment the global factor $a^{(D-1)}$.     
The equation (\ref{eq:Z}) can be written as
\begin{equation}
Z=\int [d\theta] \exp\{ -S_{gauge}(\theta)\} Z_F(\theta),         
\end{equation}
where
\begin{equation}
Z_F=\int [ d\chi^\dagger d\chi ]
            \exp\{\frac{1}{2}\sum_\ell             
                   \eta_\ell (\chi^\dagger_r U_\ell 
                  \chi_{r+a\hat{\mu}} + H.C.)         
\},
\end{equation}
where $S_{gauge}$ stands for the pure gauge 
part and $Z_F$ denotes the
fermionic partition function.  Now if we 
expand the exponential in $Z_F$ we get
\begin{eqnarray}
Z_F=\int [ d\chi^\dagger d\chi ] \prod_\ell
\exp\{\frac{1}{2}        
\eta_\ell \chi^\dagger_r U_\ell \chi_{r+a\hat{\mu}}
\}\prod_\ell
\exp\{\frac{1}{2}       
\eta_\ell \chi_{r+a\hat{\mu}}^\dagger U_\ell^\dagger \chi_r
\}    \nonumber \\
=\int [ d\chi^\dagger d\chi ] \prod_\ell
(1+\frac{1}{2}         
\eta_\ell \chi^\dagger_r U_\ell \chi_{r+a\hat{\mu}})
         \prod_\ell
(1+\frac{1}{2}         
\eta_\ell \chi_{r+a\hat{\mu}}^\dagger U_\ell^\dagger \chi_r),
\label{eq:ZF}
\end{eqnarray}
i.e. in the above product we 
have to consider each link and its opposite.

        Let us recall the rules of Grassmann variables calculus:
\begin{eqnarray}
\{ \chi_r, \chi_s\}=\{ \chi^\dagger_r, \chi^\dagger_s\}=
\{ \chi_r, \chi^\dagger_s\}=0,\\
\int d\chi_r =\int d\chi^\dagger_r = 0, \\
\int d\chi_r \chi_r =\int d\chi^\dagger_r \chi^\dagger_r = 1.
\end{eqnarray}
        
        Therefore, when we expand the 
products in (\ref{eq:ZF}) and
the Grassmann variables are integrated out, the only
non-vanishing contributions arise from 
these terms with $\chi^\dagger_r$ and
$\chi_r$ appearing one and only one time for every site $r$.  
In other words, the integration of the Grassmann variables 
produce products of $U_\ell$'s along closed paths i.e.
Wilson loops.
We denote by  ${\cal F}$ a generic configuration of 
multicomponent paths. ${\cal F}$ is specified by a set of oriented 
links verifying the rule that they
enter and leave one and only one time each lattice site $r$.  
This self-avoiding character is the
geometric expression of the Pauli exclusion principle.

We distinguish two parts in ${\cal F}$: the set 
of connected closed paths where a link is never run in 
more than once, ${\cal F}^c$ and the set of isolated
links traversed in both opposite directions or ``null''
links $\tilde{{\cal F}}$. Then 
${\cal F}={\cal F}^c\cup\tilde{{\cal F}}$. The number of 
connected closed components
of each kind is called $N_{{\cal F}^c}$ 
and $N_{\tilde{{\cal F}}}$.
In FIG.2 we show a two-dimensional sketch on a $2\times2\times3$ 
lattice of a possible
configuration ${\cal F}$ consisting in one fermionic loop 
and two ``null'' links.

\begin{center}
\begin{figure}[t]
\hskip 1cm \psfig{figure=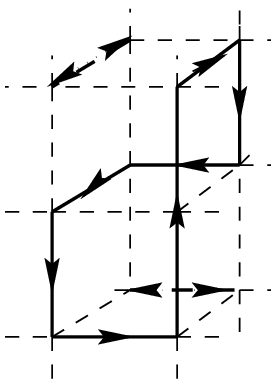,height=3.9cm}
\caption{A  
possible configuration of self-avoiding paths
in a $2\times2\times3$ lattice. 
Fermionic loops are represented by filled bold
lines and ``null'' links by 
dashed bold lines.}
\label{fig2}
\end{figure}
\end{center}

The Grassmann integration over each element of 
${\cal F}^c$ or $\tilde{{\cal F}}$ gives a $-1$.
Thus, after integrating the fermion fields, the fermionic  
path integral becomes
\begin{equation}
Z_F=\sum_{\cal F} 
(-1)^{N_{{\cal F}^c}}(-1)^{N_{\tilde{{\cal F}}}}
\prod_{l\in {\cal F}^c}
\eta_\ell U_\ell,
\end{equation}
where $V$ is the total number of lattice sites and we
used that the terms $\eta_\ell U_\ell$ for $\ell$ in the part 
$\tilde{{\cal F}}$ of ${\cal F}$ cancel out. 
It is easy to check that the number $N_{\tilde{{\cal F}}}$
of ``null'' links in a given ${\cal F}$
is connected with the number $N_{{\cal F}^c}$ of fermionic
loops as follows\footnote{This is true 
for a lattice with an even number of sites
(a lattice with an odd number of sites can not be
``filled'' with ${\cal F}$ paths).}
\begin{equation}
N_{\tilde{{\cal F}}}=\frac{V - L_{{\cal F}^c}}{2},
\label{eq:Ns}
\end{equation}
where $L_{{\cal F}^c}$ is the number of links in ${\cal F}^c$
(the length). Thus, up to a global sign, we have
\begin{equation}
(-1)^{N_{\tilde{{\cal F}}}}=(-1)^{-\frac{L_{{\cal F}^c}}{2}}.
\label{eq:NF}
\end{equation}

One can define the $\eta_{{\cal F}^c}$ for a fermionic 
loop ${\cal F}^c$:
\begin{equation}
\eta_{{\cal F}^c}=\prod_{l \in {\cal F}^c} \eta_\ell 
= \prod_{p \in S_{{\cal F}^c}} \eta_p =(-1)^{A_{{\cal F}^c}},
\label{eq:eta}
\end{equation}
where $A_{{\cal F}^c}$ is the number of plaquettes which make 
up any surface $S_{{\cal F}^c}$ 
enclosed by ${\cal F}^c$ (different 
choices of $S_{{\cal F}^c}$$S_{{\cal F}^c}$
differ by a even number of plaquettes so the sign is well defined
and one can choose the $A_{{\cal F}^c}$ as the minimal
area bounded by ${\cal F}^c$ ).
Neglecting a global sign we get
\begin{equation}
Z_F=\sum_{\cal F}^c 
\sigma_{\cal F}\prod_{l\in {\cal F}^c}U_\ell,
\label{eq:ZFfin}
\end{equation}
where
\begin{equation}
\sigma_{\cal F}=(-1)^{N_{{\cal F}^c}-
\frac{L_{{\cal F}^c}}{2}+A_{{\cal F}^c}}.
\label{eq:sigma}
\end{equation}
Let us analyze a little closer the sign $\sigma_{\cal F}$.
For $D=2$ we show in the APPENDIX I that all the 
non-vanishing contributions have $\sigma_{\cal F}=+1$.
In more than two space-time dimensions there are non-null
contributions ${\cal F}$ to $Z_F$ with both signs.
The reason is that 
connected fermionic loops enclosing odd numbers of vertices
do not necessarily imply any more a null contribution. 
Different examples of a simple fermionic loop contributing with
For instance, in case (a) the enclosed area 
$A_{{\cal F}^c}$ is 4, the lenght  
$L_{{\cal F}^c}$ is 8.
Therefore, the fermionic loop (a) has
$N_{{\cal F}^c}-
\frac{L_{{\cal F}^c}}{2}+A_{{\cal F}^c}=1-4+4=1$
, and then $\sigma_{\cal F}=-1$.
%( 1-8+10=3 ), and then $\sigma_{\cal F}=-1$ in both cases.
%It is seen that both signs do not appear symmetrically.

\begin{center}
\begin{figure}[t]
\hskip 1cm \psfig{figure=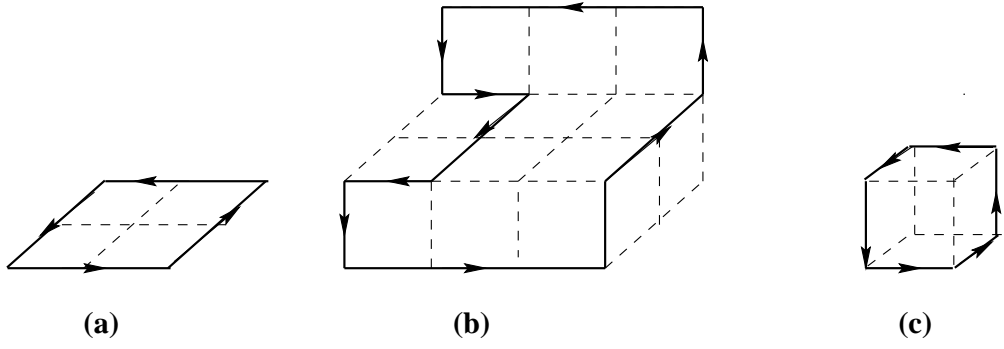,height=4.5cm}
\caption{Two examples of fermionic loops 
contributing with a -1.}
\label{fig3}
\end{figure}
\end{center}

        Coming back to the total path integral (\ref{eq:Z}) 
we can write it as:
\begin{equation}
Z=\int [ d\theta ] \sum_{n}
\sum_{\cal F}
\sigma_{\cal F}^c
(\prod_{\ell\in {\cal F}^c}U_\ell)
\exp\{ -\frac{\beta}{2} \|\nabla \theta +2\pi n\|^2\}
\label{eq:diome}
\end{equation}

We can express the fermionic paths ${\cal F}^c$ in terms 
of integer 1-forms
--attached to the links-- $f$ with three possible values: 0 and
$\pm$ 1 with the constraint that they are 
non self-crossing and closed $\partial f = 0$   where
$\partial$ is the boundary operator adjoint of $\nabla$ 
which maps $k$-forms into $(k-1)$-forms. Both operators verify
the integration by parts rule
\begin{equation}
<\partial g, h>=<g,\nabla h>,
\label{eq:b-cb}
\end{equation}
where $g$ and $h$ are respectively $k$ and $k-1$ arbitrary forms.
The factor $\prod_{l\in {\cal F}^c}U_\ell$ in (\ref{eq:diome})
is nothing but the product of Wilson loops along the f-loops i.e.
$\exp\{ i\sum_{\ell\in {\cal F}^c}\theta_\ell f_\ell\}$.
The $\tilde{{\cal F}}$ can be expressed
by means of functions attached to links
$\tilde{f}$ with value 0 or 1 and defined
over $f^{-1}(0)$, the ``nucleus'' of $f$.
In terms of the $f$ and $\tilde{f}$ we get:
\begin{equation}
Z=
\int [d\theta]\sum_{n}
\sum_{f} \sum_{\tilde{f}}  
\, \exp\{
-\frac{\beta}{2} \|\nabla \theta +2\pi n\|^2 +i<\theta,f>
\}.
\end{equation}
If we use the Poisson summation formula $\sum_n g(n) 
= \sum_{n'} \int_{-\infty}^{\infty} dB g(B) e^{2\pi iBn'}$ 
--where $n$ and $n'$ are
integer 2-forms and $B$ is a real 2-form-- we get
$$
Z=
\int [d\theta]\sum_{n}
\sum_{f} \sum_{\tilde{f}} 
\sigma(f)  \times 
$$
\begin{equation}
\int_{-\infty}^{\infty} [dB]
\exp\{
-\frac{\beta}{2} \|\nabla \theta +2\pi B\|^2 +
i<\theta,f>+2\pi i<B,n>
\}.
\end{equation}
Performing the displacement $B \rightarrow B-\nabla 
\theta /2\pi$ and integrating in $B$
$$
Z=
\int [d\theta]\sum_{n}
\sum_{f} \sum_{\tilde{f}} 
\sigma(f) \, 
\exp\{
-\frac{1}{2\beta} \|n\|^2 + i<\theta,f-\partial n>
\}
$$
$$
=\sum_{n}\sum_{f} \sum_{\tilde{f}} 
\sigma(f) \, 
\exp\{
-\frac{1}{2\beta} \|n\|^2
\}
\delta (f-\partial n)
$$
\begin{equation}
=\sum_{n} \sum_{\tilde{f}}
\sigma(\partial n)
\exp\{
-\frac{1}{2\beta} \|n\|^2
\},
\label{eq:Zs}
\end{equation}
where we have used equation (\ref{eq:b-cb}) in order to
transform $<\nabla \theta, n>$ into $<\theta,\partial n>$
whose integration produced the Dirac's delta $\delta(f-
\partial n)$. 
Equation (\ref{eq:Zs}) is a geometrical expression of the path
integral of  lattice $QED$ with staggered fermions
in terms of surfaces with self-avoiding boundaries.

	Now, let us return to our goal, namely to set up the 
worldsheet Lagrangian formulation corresponding to the 
P-representation of sect. 2. The path integral (\ref{eq:Zs}) 
includes worldsheets of paths with ends of any parity. 
Therefore, obviously, it does not produce the Hamiltonian 
(\ref{eq:H}) by the transfer matrix method. 

As a matter of fact, the transfer matrix for staggered fermions is
a general and interesting problem on which not much has been done.
Unlike the Wilson fermions which come out rather nicely from the
transfer matrix \cite{cr}- \cite{lu}, the staggered fermions
present some troubles.  For the spatial part of the action the
derivation goes in parallel with that for the Wilson theory.  The
interesting question is how does the temporal part work.  This
problem was discussed in the work of Saratchandra et all
\cite{stw}, in this paper was analyzed the transfer matrix
formalism for relating the Kogut-Susskind Hamiltonian and the
Euclidean action for staggered fermion fields.  The authors pointed
out that the ordinary case, in which the one-component fermions
$\chi(x)$ and $\chi^\dagger(x)$ both live on all sites, exhibits a
doubling of the fermion species with respect to the hamiltonian
formalism (four flavours instead of two flavours).  Additionally,
the corresponding transfer matrix is hermitean but not positive
definite.  They explored the alternative of considering $\chi$ and
$\chi^\dagger$ on alternate sites, although this procedure avoids
the excess of flavours and gives a positive transfer matrix it is
not free from complications.

Although the partition function (\ref{eq:Zs}) is
not directly connected with the Hamiltonian (\ref{eq:H})
via the transfer matrix, it is closely related with it (as
we will show) and 
provides us with a guide to guess the genuine partition 
function for the P-representation $Z_P$.
With this aim we consider the restriction on 
the set of the surfaces, to the subset of  
the surfaces such that they are worldsheets of the $P$ paths.
This is equivalent to require that 
when intersected with a time 
$t$ = constant plane they give the paths of $\{ | P > \}$ 
i.e. paths with ends of opposite parity and oriented from even 
sites to odd sites.  
 We get a link of $P_t$ for every 
plaquette of the surface which connects the 
slice $t$ with the $t+a_0$.
As an illustration, in FIG.4 we show the paths of the 
P-representation we get 
from the minimal surface enclosed by the configuration of 
self-avoiding ${\cal F}^c$ paths depicted in FIG. 2.
\begin{center}
\begin{figure}[t]
\hskip 1cm \psfig{figure=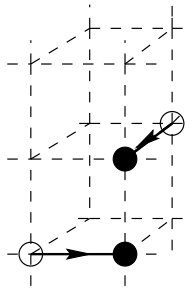,height=3.9cm}
\caption{The P-paths (thick lines) for the different
times corresponding to the surfaces enclosed by the 
configuration of self-avoiding paths of FIG.2.}
\label{fig4}
\end{figure}
\end{center}
The imposed constraint forbids fermionic loops such that
the one of FIG.3-(c) which produces at the first temporal slice
a path connecting vertices with the same parity. 
The first configuration with negative Boltzman factor
which appears in the partition function, i.e. that of
lower action, is the spatial
square of side 2 depicted in FIG.3-(a) with 
area $A_{{\cal F}^c}=4$.

Thus, we propose the following worldsheet partition function:
\begin{equation}
Z_P= 
\sum_{S} \sigma_h(S)\exp\{
-\frac{1}{2\beta} \sum_{p \in S} n_p^2
\},
\label{eq:ZP0}
\end{equation}
where $S$ runs over worldsheets
of P-paths and $\sigma_h(S)$ is a sign defined:
\begin{equation}
\sigma_h (S) = \prod_t(-1)^{A_t}\nu(P_{t-1}\partial S_t)\nu(P_{t})
\label{eq:sigma_h}
\end{equation}
where $S_t$ are the surfaces given by the spatial plaquettes
of $S$ at section $t$, $A_t$ is the number of plaquettes
in $S_t$ and the function $\nu$ gives a sign defined by
\begin{equation}
\nu( P) = \prod_\alpha(-1)^{\frac{|P_\alpha|-1}{2}}
\prod_\beta(-1)^{\frac{|Q_\beta|}{2}-1}
\end{equation}
($P=\{ P_\alpha, Q_\beta\}$ being
$P_\alpha$ and $Q_\beta$ the open and closed component
paths respectively. See APPENDIX II for notation and discussion.) 
Additionally, note that we have eliminated the sum over 
the ``null'' links.
The rationale for this is that the ``null'' links do not play any
role in the P-representation. In the next section we are going to
show  explicitly that $Z_P$ gives rise to the Hamiltonian 
(\ref{eq:H}) via the transfer matrix procedure.

\vspace{1cm}

\section{Hamiltonian obtained via the transfer Matrix method}

\vspace{3mm}

By means of the transfer matrix 
method let us show that we
re-obtain the Hamiltonian (\ref{eq:H}) from 
the path integral $Z_P$.
As we wish to consider the continuous time limit of
the previous lattice Euclidean space-time theory, we
introduce a different lattice spacing $a_0$ 
for the time direction.
The couplings on timelike and spacelike
plaquettes are no longer equal in the action
i.e. we have two coupling constants: $\beta_0$ and 
$\beta_s$. The temporal coupling constant $\beta_0$ 
decreases with $a_0$ whilst the spatial coupling constant
$\beta_s$ increases with $a_0$.
So far we have neglected a factor ${\frac{a^{(D-1)}}{2}}^V$ 
in the path integral. Taking into account the fact
that the lattice has a different temporal separation 
$a_0\neq a$ we get a relative factor of  
${(\frac{a_0}{a})}^{|f_\ell |}$ 
for each spatial link of ${\cal F}^c$. 
\begin{equation}
Z_P= 
\sum_{S} a_0^{|\partial n|_{sp}}\sigma_h(S)\exp\{
-\frac{1}{2\beta} \sum_{p \in S} n_p^2
\},
\label{eq:ZP0T}
\end{equation}
where $|\partial n|_{sp}$ denotes the number of spatial links
in $\partial n$. To factorize $Z_P$ to fixed time contributions
we consider the spatial plaquettes of $S$ that for each $t$
define the spatial surfaces $S_t$ and the temporal plaquettes
that define the spatial paths $P_t$ and
we write
\begin{equation}
-\frac{1}{2\beta}\| n\|^2 = 
-\frac{1}{2\beta_s}\sum_t \| n_t\|^2-\frac{1}{2\beta_0}
\sum_t\| P_t\|^2
\end{equation}
and
\begin{equation}
\sigma_h (S) = \prod_t \sigma^t(P_{t-1},P_{t},S_t)
\end{equation}
\begin{equation}
\sigma^t(P_{t-1},P_{t},S_t)=(-1)^{A_t}\nu(P_{t-1}\partial S_t)
\nu(P_{t})
\end{equation}

To write the operator $\hat{T}$ which connects the ket
$|P_{t-1}>$ with the ket $|P_t>$ we begin by decomposing
the sum over world sheets $\sum_{S}$ in (\ref{eq:ZP0T}) 
into 2 sums: one over the temporal parts, the $P_t$, and 
one over spatial parts $S_t$ i.e.
$$
Z_P= \sum_{\{P_t\}}\, \prod_t <\, P_{t}| \hat{T} | \, P_{t-1} >
$$
\begin{equation}
<\, P'| \hat{T} | \, P >=\sum_{S_t}\,
a_0^{|\partial S_tP'\bar{P}|_{sp}} \sigma^t
\exp\{
-\frac{1}{2\beta_0} \| P\|^2
-\frac{1}{2\beta_s} \| n_t\|^2
\}
\label{eq:PTP}
\end{equation}

Taking into acount the relation which connects $\hat{T}$ 
with $\hat{H}$ when $a_0$ is small:
\begin{equation}
\hat{T}\propto e^{-a_0\hat{H} + O(a_0^2)},
\label{eq:T-H}
\end{equation}
we find  $\hat{H}$ considering this limit
$$
<\, P'| \hat{T} | \, P >\approx \delta_{P,P'}\,
-a_0<\, P'| \hat{H} | \, P >
$$
The dominant contributions in this limit are the
cases where $P'$ is equal to $P$ or differs of it
by one link $\ell$ or by one plaquette $p$. The power of $a_0$
in (\ref{eq:PTP})
forces $S_t=0$ in the first two cases and
$S_t\equiv p$ in the third. 
\begin{equation}
<\, P| \hat{T} | \, P >\approx 1 \approx
\exp\{
-\frac{1}{2\beta_0} \| P\|^2
\}
\label{eq:PP}
\end{equation}
\begin{equation}
<\, P\cdot p| \hat{T} | \, P >\approx -a_0
<\, P\cdot p| \hat{H} | \, P > \approx
-\exp\{-\frac{1}{2\beta_s} \}
\label{eq:PpP}
\end{equation}
\begin{equation}
<\, P\cdot \ell| \hat{T} | \, P >\approx 
-a_0<\, P\cdot \ell| \hat{H} | \, P > \approx
a_0\sigma^t
\label{eq:PlP}
\end{equation}
To obtain a proper continuum time limit we should take 
\begin{eqnarray}
\beta_0= \frac{a}{g^2a_0} \\
\beta_s= \frac{1}{2} 
\frac{1}{\mbox{ln} (2g^2 a/a_0)} \, ,
\label{eq:betas}
\end{eqnarray}
where $a$ continues to denote the spacelike spacing.
The values of $\sigma^t$ are $-1$ when adding a plaquette
(\ref{eq:PpP})
and depend on the way we do it when adding a link (\ref{eq:PlP}).
An analisis of the possible cases shows that $\sigma^t$
corresponds to the sign $\epsilon$ in (\ref{eq:Wi}).

Then,
\begin{equation}
a_0\hat{H}| \, P > =
 \frac{g^2}{2a} \sum_\ell N_\ell^2(P)\mid P >+
 \frac{1}{2ag^2} \sum_{p}(\mid P \cdot p>+
\mid P \cdot \overline{p}>)
 -\frac{1}{2a}\sum_{\ell}\epsilon (P,\ell)\mid P\cdot \ell>
\label{eq:HPT}
\end{equation}
so we recover the Hamiltonian (\ref{eq:H}).
confirming that (\ref{eq:ZP0}) is 
the expression of the partition 
function of compact electrodynamics  
in the P-representation.

In FIG.  5 we show an scheme which summarizes the different
actions/hamiltonians for QED with staggered fermions, and their
connections.  The surface action $S_{\mbox{surf}}$ of
(\ref{eq:Zs}), which was obtained from the Kogut-Susskind action
$S_{KS}$ by integrating over the gauge and fermion fields, does not
produce the hamiltonian $H_P$ which in fact is connected with the
$S_P$ of (\ref{eq:ZP0}).  This is because $S_P$ is free from the
known problem of $S_{KS}$ of having an unwanted additional doubling
compared with the hamiltonian formulation.  
On the other hand, the equivalence
between the Kogut-Susskind Hamiltonian $H_{KS}$ and the
P-Hamiltonian $H_P$ was proved in \cite{fg}.

\vspace{1cm}

\begin{center}
\begin{figure}[t]
\hskip 1cm \psfig{figure=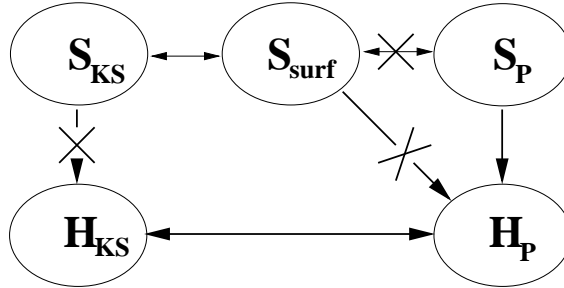,height=3.8cm}
\caption{A diagram summarizing the different 
lattice actions and Hamiltonians 
for compact QED and their connections. 
The horizontal lines denote
equivalence. The vertical lines emanating from the classical
actions to the hamiltonians represent the connection via
the transfer matrix procedure.}
\label{fig5}
\end{figure}
\end{center}

In other words, the action $S_P$ should be regarded as a different
lattice action, which produce a Hamiltonian equivalent to the
Kogut-Susskind's one.

\section{Conclusions and Final Remarks}

We propose a purely geometric action in terms of the
world-sheets of P-path configurations.
In this formulation, the connection 
between the Hamiltonian and Lagrangian of 
QED with staggered fermions is straigthforward
via the transfer matrix method.
The partition function can be written as a sum over surfaces with
border on fermionic self-avoiding loops ${\cal F}^c$ . 
Hence, the fermionic problem has been 
reduced to the task of computing quadratic areas 
enclosed by polymer-like configurations. 
The polymer representation of 
lattice fermions \cite{montvay} is often used 
in a different way to compute the fermionic
determinant. 
It is important to note that our formalism is free from
the problem 
of the additional doubling of
the fermion species due to the discretization of time.

With regard to the economy and possible advantages from the 
numerical computation point of view offered by the  
P-description, we want to emphasize two facts:
I) Concerning the gauge degrees of freedom, it only involves 
sums of gauge invariant variables i.e. no gauge redundancy.
II) It includes a subset of the configurations
which are taken into account in the path integral 
(\ref{eq:Zs}) equivalent to the one of Kogut-Susskind.
Our formulation involves a sum of configurations with
Boltzman factors of both signs. However,
the lower action configuration with negative Boltzman factor
which appears in the partition function
is the spatial square of side 2 which has 
area $A_{{\cal F}^c}=4$. 
This shows that the positive and
negative Boltzman weights are not balanced and that techniques
such that the histogram method \cite{fs88} can be applied.
Equipped with the geometrical insight provided by this 
gauge invariant representation, we are working to
design a suitable algorithm for simulating the loop 
fermionic action. 
The simplest case is $QED$ in (1+1) 
dimensions or the Schwinger model for which 
all the Boltzman factors
are positive i.e. one does not have to worry about
computing the sign $\sigma_h(S)$ . 
In addition, in a two space-time lattice there is a one 
to one correspondence between fermionic loops 
and the surfaces with border on them.
So, to evaluate the path integral, the procedure
is to generate surfaces with  
self-avoiding frontiers and with 
the constraint that they produce on each time slice  
open paths with ends of opposite parity. This was
done recently \cite{f97} by means of a Metropolis Monte Carlo
algorithm and the results are very encouraging. 

Finally, let us mention that in this paper 
we only considered the 
simpler Abelian massless theory. 
In the non-Abelian case (see ref. \cite{na} for the loop 
formulation of the pure gauge theory), where there are different
colors, the Pauli exclusion principle implies
that the maximum number of such pairs at any
site can not exceed the total number of degrees
of freedom of the quarks. Therefore,
more complicated diagrams arise. For instance,
intersecting fermionic loops at vertices
where there are more than a $q\bar{q}$ pair. 
Here the path to be followed is not unique.
It is interesting to consider the massive case too. 
In this case, in addition to the 
{\em dimers}, we would have {\em monomers} produced 
by the mass term. 
The construction of a lattice path integral
in terms of loops for {\em full} QCD is a
task that will be considered in a future work.

\vspace{5mm}

{\large \bf Acknowledgements}

\vspace{2mm}

We wish to thank helpful comments 
and remarks from Michael Creutz.

This work was supported in part by CONICYT, 
Projects No. 318 and No. 49.

\section*{Appendix I}

A general configuration ${\cal F}$ with $\sigma_{\cal F}=-1$
obviously must include at least one connected fermionic loop 
contributing with a $-1$ to $\sigma_{\cal F}$. 
We will prove here that such connected 
fermionic path necessarily
encloses an odd number of sites and thus, in $D=2$, 
it is not possible to  ``fill'' its interior region
in such a way that all the sites are ``visited'' 
( the condition required for a non-null contribution to $Z_F$).

Notice that, according to equation (\ref{eq:sigma}),
the $\sigma_{\cal F}$ associated to a configuration with
a unique connected fermionic loop is negative if
$L_{{\cal F}^c}/2+A_{{\cal F}^c}$ is an even number.

In order to prove that all the fermionic loops
of  ${\cal F}^c$ which give a $-1$ contribution 
to $\sigma_{\cal F}$ 
enclose an odd number of sites we need to express 
the number of its inner vertices as a function  
of its length and area. 
In $D=2$, 
a generic connected fermionic loop ${\cal F}^c$
enclosing an arbitrary number of vertices
can be obtained by a constructive procces begining with 
a plaquette $p$ and then generating all
the others fermionic loops by ``appending'' plaquettes
to this diagram on each of its links in
the two different ways showed in FIG. 6.

\begin{center}
\begin{figure}[t]
\hskip 1cm \psfig{figure=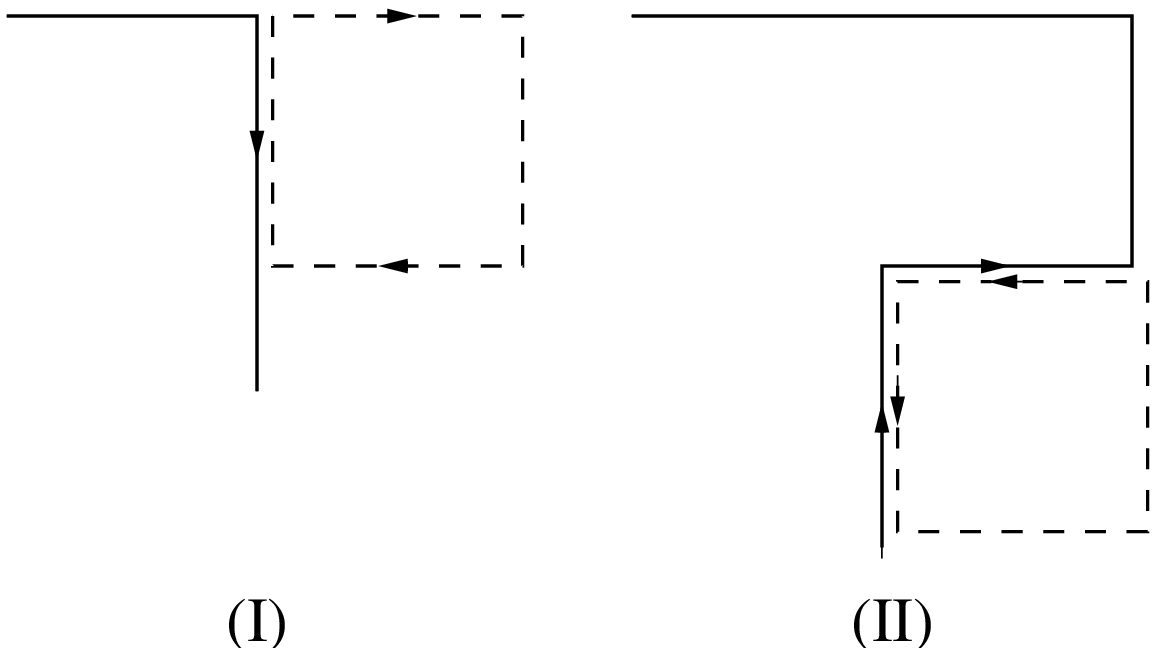,height=4cm}
\caption{}
\label{fig6}
\end{figure}
\end{center}

The variations of $N_{{\cal F}^c}$
$\frac{L_{{\cal F}^c}}{2}$ and $A_{{\cal F}^c}$ 
for the two cases illustrated in FIG. 6 can be summarized as
follows:

\begin{eqnarray}
&\mbox{I)} \;
\Delta A_{{\cal F}^c}= 1 \;\;\; 
\Delta L_{{\cal F}^c}= 2, \nonumber \\
&\mbox{II)} \;
\Delta A_{{\cal F}^c}=1 \;\;\; 
\Delta L_{{\cal F}^c}=0. \nonumber \\
\end{eqnarray}

From the above relations it is easy to check 
\begin{eqnarray}
&\Delta A_{{\cal F}^c}=N_{p_I}+N_{p_{II}},\\
&\Delta L_{{\cal F}^c}=2N_{p_I}. \nonumber
\end{eqnarray}
where $N_{p_I}$ and $N_{p_{II}}$
denote the number of plaquettes of type $I$ and $II$ 
respectively. It is also easy to see that 
the number of inner vertices $I_{{\cal F}^c_k}$
to a connected loop 
is equal to $N_{p_{II}}$. Therefore, since
$A_{{\cal F}^c}= 1+\Delta A_{{\cal F}^c}$ and
$L_{{\cal F}^c}= 4+\Delta L_{{\cal F}^c}$,

\begin{equation}
I_{{\cal F}_k^c}= 1-\frac{L_{{\cal F}^c}}{2}+A_{{\cal F}^c}.
\label{eq:inner}
\end{equation}

From equations (\ref{eq:sigma}) and (\ref{eq:inner})
we conclude that 
the contribution of ${\cal F}_k^c$ to
$\sigma_{\cal F}$, $\sigma_{{\cal F}_k^c}$ is $-1$
if and only if $I_{{\cal F}_k^c}$ is an odd number.

The same procedure extends to higher dimensions.
(\ref{eq:inner}) is true if the surface is built
through steps of type $I$, $II$ and $III$ where
$III$ means adding a plaquette making contact in
three links of the border. If the surface intersects
itself, inner vertices belonging to the intersection
lines are counted with the corresponding multiplicity. 

\section*{Appendix II}

\subsection{Structure of surfaces}

Consider a surface $S$ contributing to the partition function.
Its border is $C=\partial S$, a simple loop without intersections
(not necessarily connected). This surface carries a sign:
\begin{equation}
\sigma(\partial S) =(-1)^{N-\frac{L}{2}+A}
\end{equation}
where:
\begin{eqnarray*}
N =  \mbox{Number of connected components in } C \\
L =  \mbox{Number of links in } C \\
A = \mbox{Number of plaquettes in } S
\end{eqnarray*}
The sign of $S$ depends only on $\partial S$ since two
surfaces with the same border differ by an even
number of plaquettes.

We divide $S$ in spatial and temporal plaquettes:
$S=\{ S_t, P_t\}$ where
%\begin{equation}
%\begin{array}{l}
\begin{eqnarray*}
S_t = \mbox{ Spatial plaquettes of $S$ in the section $t$}\\
P_t = \mbox{ Spatial links in the section $t$
 corresponding} \\\mbox{ to temporal 
 plaquettes of $S$ between $t$ and $t+1$}
\end{eqnarray*}

Now we have at each spatial section the paths $\partial S_t$. Part
of these paths propagates to $t+1$, part comes from $t-1$ and part
does not propagate and belongs to $C$.
\begin{eqnarray*}
B_t = \mbox{ Part of $\partial S_t$ that does not propagate}\\
R_t^+ = \mbox{ Part of $\partial S_t$ that propagates to $t+1$}\\
R_t^- = \mbox{ Part of $\partial S_t$ that comes from $t-1$}
\end{eqnarray*}

\begin{equation}
\partial S_t = B_t\cdot  R_t^+\cdot  R_t^- 
\end{equation}                                                      

Let us localize now the links of $C$. Its temporals links 
correspod to $\partial P_t$. Its espatial links belong
to espatials plaquettes, in which case they belong to
$A_t$, or to temporal plaquettes that can move up or down.
\begin{eqnarray*}
K_t^+ = \mbox{ Spatial part of $C$ in section $t$ 
that propagates to $t+1$}\\
K_t^- = \mbox{ Spatial part of $C$ in section $t$ 
that comes from $t-1$}
\end{eqnarray*}

To keep track of all links in $S$ (those that
belong to any plaquette in $S$)
we consider those joining 
two spatial plaquettes or two temporal plaquettes.
The first ones are irrelevant to the sign.
The second ones can join two temporal plaquettes
in the same section (irrelevant) or in correlative
sections:
\begin{eqnarray*}
\Pi_t = \mbox{ Spatial links of $S$ in section 
$t$ that propagate to $t+1$}\\
\mbox{ and come from $t-1$}
\end{eqnarray*}

To sum up (keys denote the set for all $t$ and
parenteses for a fixed $t$)
\begin{eqnarray*}
S=\{ S_t, P_t\}\\
\partial S_t = B_t R_t^+ R_t^- \\
C_t=( B_t, K_t^+, K_t^-)\\
P_t = ( \Pi_t, K_t^+, \bar{R}_t^+)\\
C=\{ C_t, \partial P_t\}\\
\end{eqnarray*}

Note that links in $B_t, K_t^+, K_t^-$ are incompatible
among them while $\Pi_t, R_t^+, R_t^-$ can share
parts among them and with the former. 

There is a balance of creation and annihilation
\begin{equation}
\sum_t (|K_t^+|+|R_t^+|) = \sum_t (|K_t^-|+|R_t^-|).
\end{equation}

And what leaves at $t-1$ is what arribes to $t$
\begin{equation}
P_{t-1} = ( \Pi_{t-1}, K_{t-1}^+, \bar{R}_{t-1}^+) = 
( \Pi_t, \bar{K}_t^-, R_t^-).
\end{equation}

It also verifies
\begin{equation}
(-1)^{A-\frac{L}{2}}=(-1)^{\sum_tA_t-\frac{1}{2}
\sum_t(|B_t|+|K_t^+|+|K_t^-|)}
\end{equation}

\subsection{Localization of the sign}

The size of
an objet $X$ is $|X| =$ number of links or plaquettes
(depending if it is a path or a surface) taking into
account its multiplicity.
A path is fermionic if it consists of closed single lines
and open single lines from even sites to odd sites.
For these paths, $P=\{ P_\alpha, Q_\beta\}$ where
$\alpha$ labels the open components and $\beta$ the closed ones,
we define the function
\begin{equation}
\nu( P) = \prod_\alpha(-1)^{\frac{|P_\alpha|-1}{2}}
\prod_\beta(-1)^{\frac{|Q_\beta|}{2}-1}
\end{equation}
The meaning of this sign is that it is the sign we
get when we create a single 
line of lengh $L$ using only the  $\hat{\Phi}_\ell$ operators.
There are $(L-1)/2$ joining negative
actions  if the line is open and $L/2-1$ if the line is closed.

The sign in (\ref{eq:ZP0}) is defined
\begin{equation}
\sigma_h (S) = \prod_t(-1)^{A_t}
\nu(P_{t-1}\cdot \partial S_t)\nu(P_{t})
\end{equation}
and it can be written
\begin{equation}
\sigma_h (S) = \prod_t \sigma^t(P_{t-1},P_{t},S_t)
\end{equation}
\begin{equation}
\sigma^t(P_{t-1},P_{t},S_t)=(-1)^{A_t}
\nu(P_{t-1}\cdot \partial S_t)\nu(P_{t})
\end{equation}

\end{document}